# SNOWMASS WHITE PAPER:
## SLHC Endcap 1.4≤η≤4 Hadron Optical Calorimetry Upgrades in CMS with Applications to NLC/T-LEP, Intensity Frontier, and Beyond

Burak Bilki[2], Yasar Onel[2], David R Winn[1*], Taylan Yetkin[2]

Fairfield University[1], University of Iowa[2]

*Corresponding: winn@fairfield.edu
1+203.984.3993

**Abstract:** Radiation damage in the plastic scintillator and/or readout WLS fibers in the HE endcap calorimeter (1.4<η≤3 ) in the CMS experiment at LHC and SLHC will require remediation after ~2018. We describe one alternative using the existing brass absorber in the Endcap calorimeter, to replace the plastic scintillator tiles with $BaF_2$ tiles, or quartz tiles coated with thin(1-5µm) films of radiation-hard pTerphenyl(pTP) or the fast phosphor ZnO:Ga. These tiles would be read-out by easily replaceable arrays of straight, parallel WLS fibers coupled to clear plastic-cladded quartz fibers of proven radiation resistance. We describe a second alternative with a new absorber matrix extending to 1.4≤η≤4 in a novel Analog Particle Flow Cerenkov Compensated Calorimeter, using a dual readout of quartz tiles and scintillating (plastic, $BaF_2$, or pTP/ ZnO:Ga thin film coated quartz, or liquid scintillator) tiles, also using easily replaceable arrays of parallel WLS fibers coupled to clear quartz transmitting fibers for readout. An Analog Particle Flow Scintillator-Cerenkov Compensated Calorimeter has application in NLC/T-LEP detectors and Intensity Frontier detectors.

**Introduction:** The forward hadron calorimeter regions η≥1.4 in CMS/LHC must survive large radiation doses in the first 1-2 interaction lengths, and have limited redundant information beyond η≥3, with no tracker e-m calorimeter or muon system. For CMS we have proposed a novel exceptionally rad-hard and fast (sub-25ns) electromagnetic front end calorimeter to help protect the present Forward Calorimeter HF (3≤η≤5) to add 1-2 λ thickness, and to mitigate pile-up in the present HF. This new Secondary Emission (SE) Calorimeter technique uses secondary emission dynode planes as the active medium to measure the energy of showers in absorption calorimeters, and is described in an accompanying Snowmass White Paper[1]. A Forward Lepton-Photon System fproposed for CMS is described in another accompanying Snowmass White Paper[2], that also adds superferric muon toroids outside the present Forward Calorimeter HF (to measure muons in CMS from 2.4≤η≤5, for direct H->µµ, WW asymmetry, and BSM physics.

Accompanying improvements to the highest forward region, the scintillating tile/WLS endcap HE hadron calorimeter[3] (1.4<η≤3) in CMS will require modification or replacement to survive 10 years of LHC and into SLHC. Other proposals so far include: a) retaining the present absorber and replacing the present damaging plastic scintillator tiles with new tiles with a larger number of WLS fibers so that light is collected over smaller distances. The new tiles may be green emitting with green to red WLS shifting which may be more radiation resistant; b) liquid scintillator tiles; c) replacing the existing absorber with and absorber extending to 1.4<η≤4. These include Digital/Particle Flow Calorimetry[4], as pioneered by the CALICE collaboration[5], which may use RPC[6] for example, GEMs or other pixillated pad readouts.

## 1) Rad-Hard Scintillating Tiles in the Existing HE Endcap Calorimeter

For rad-hard upgrades of the existing HE Endcap Calorimeter in CMS, quartz tiles coated with pTP or ZnO:Ga scintillator, and/or $BaF_2$ tiles(suggested by H. Newman)[7], read out by replaceable WLS fibers are proposed. These new tiles may be placed mainly in the first ~2$\lambda$ and $\eta>2$ of HE, or all of HE.

$BaF_2$ has high light output, speed in the UV (0.6ns), and good radiation resistance for the HE region, and has been extensively studied[8].

Many recent studies 2008-11[9,10,11,12] have demonstrated pTP coated quartz tiles with WLS readout for the HE calorimeter are adequate (hadronic energy resolution with ~1 $\lambda$ Fe sampling: 210%/$\sqrt{E}$, constant term 8.8%) as a radiation hard substitute for the plastic scintillator tiles presently used in HE, and may prove less expensive than $BaF_2$. Hadronic energy was integrated over less than 20 ns in test beams, and the light yield with 4-5 µm thick films of of pTP or 2-3 µm ZnO:Ga thick films of is predicted to be 5-6x Cerenkov light, based on data with 2 µm films, as discussed below.

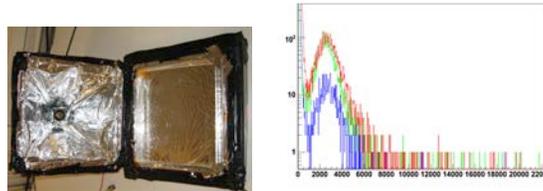

*Figs 1: (L)* A 15x15 cm, 5 mm quartz plate coated with ZnO:Ga with an aluminized mylar "pyramid" air-coupling the light emerging from the coating to a PMT at the truncated apex of the pyramid. pTP coated plates and clean quartz plates were similarly mounted. *(R)* Muon MIPs produced single electron peaks in plain quartz(blue) and in 2µm pTP film-coated(red) and 0.3 µm film-coated ZnO:Ga(green) quartz plates. Remarkably, the singles rate was x4.5-5 times higher in the scintillator coated plates, despite how very thin the scintillating/shifting films are. The means of the MIP signals were 140, 448, and 530 pC for quartz, ZnO:Ga and pTP films respectively, factors of 3.2-3.8 in light yield with the very thin scintillating/shifting films over bare quartz, whose mean and singles peak were similar.

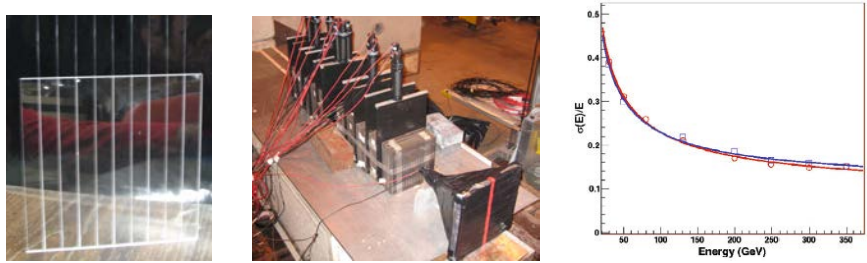

*Figs 2:* (L) A quartz plate coated with pTP, readout by parallel WLS fibers; ZnO:Ga filmed quartz or $BaF_2$ plates to be similar. (M) A photograph of the CERN 10 $\lambda$ Fe Test Beam Calorimeter with pTP 2 µm film coated Quartz Plates; a Dual Readout $BaF_2$ scintillator/Quartz Plate Calorimeter would be similar. (R) $\sigma E/E$ vs E for a quartz/pTP test calorimeter 10 $\lambda$ x 1 $\lambda$ wide.

The readout of these new tiles would use multiple parallel WLS fibers (similar to the figure above), coupled at the edge of the tiles to larger diameter quartz-cladded with plastic fibers (10% loss at 1 Gigarad). The readout would be modified so that the multiple straight-through WLS fibers are designed with mechanics which allow just the WLS fibers be easily replaced by removal of the WLS coupled to clear fibers, at periodic downtimes, similarly in concepts to the mechanisms that allow radiosource wires to be inserted and removed.

## 2) HE Replacement to η~4: Analog Particle Flow Calorimetry via Dual-Readout Scintillator/Quartz Tiles

If the HE absorber is replaced and extended to η~4, then the possibility of digital (with RPC or GEM readouts) or analog Particle Flow Calorimetry is an excellent candidate. Another candidate is Cerenkov compensated hadron calorimetry, first quantitatively proposed in MC studies[13]. Subsequently, it has been demonstrated by the TTU group lead by R.Wigmans & collaborators in the DREAM[14] parallel scintillating fiber/quartz fiber calorimeter configuration, and has proven improved energy resolution. However, despite the benefits, the dual readout with parallel fibers is unwieldly in the HE (and many NLC) application. We propose to extend that work of dual readout benefits to plate/tile calorimeter configurations, using alternating back-back layers of scintillator tiles and quartz tiles, with separate replaceable WLS readouts, and with more radiation-hard scintillator tiles. We further propose to extend such dual readout compensation to Analog Particle Flow (PF) Calorimeters. An Analog PF Calorimeter using fine-grained plastic scintillator tiles with WLS readout have been beautifully realized by the CALICE Collaboration[15]; we borrow heavily from their work, with the following changes/additions: Adding clear quartz (or aerogel) plates serially to scintillating plates for Cerenkov compensation – dual readout; 2) Substituting more rad-hard scintillators, such as $BaF_2$ tiles, pTP or ZnO:Ga film coated quartz tiles (as the above section) or liquid scintillator tiles; and 3) Straight replaceable arrays of WLS fibers coupled to quartz fiber readouts. This concept is better suited to an HE replacement as discussed below:

*1) Particle Flow Compatibilities:*
Parallel fiber geometries are incompatible with PF calorimeter configurations, where small volumes of the calorimeter are read-out. Scintillating tiles & clear tiles of near-arbitrarily small sizes can be read-out via WLS fibers coupled to clear transmission fibers. This brings the virtue of Cerenkov compensation to enhance analog particle flow. We propose a staged plan where ΔηxΔϕ towers at first are similar to existing jet cones, and have relatively minimal longitudinal segmentation as in the existing HE/HB. However the plates would be composed of several tiles readout out by separate WLS fibers, ganged together to form partial longitudinal segmentation compartments, but which in future upgrades, by separating the readout fibers and adding photodetectors, could approach particle flow transverse and longitudinal segmentation. The WLS readout fibers would be as described above for HE plate replacement.

*2) Pointing Towers:* Fiber calorimeters are cumbersome to be made pointing at a "target"=IP, over a large enough range of solid angles in ΔηxΔϕ towers to remain quasi-homogeneous, whereas plates can be formed in arbitrary shapes and sizes, positioned to be readout as constant projective solid angle towers growing in transverse area from front to back. Pointing ΔηxΔϕ physics towers are difficult with parallel fiber geometry, at the angles varying over *1.5<η<4* (25°-2°). Already in HF, the η =3 towers are at ~5° to the pointing direction. Various solutions proposed by advocates of the parallel fiber dual readout show this difficulty.

Proposed solutions to the pointing problem are fiber channels arranged along radii pointing to the IP – difficult mechanically, but possible in principle. The packing fraction is thus a function of distance into the calorimeter (less at the outer surface), and introduces a constant term, obviating some of the energy resolution advantages. A further modification on these lines is to make tapered radially pointing (i.e. pointing to the IP) grooves, and produce tapered fibers. This is not very practical, and also introduces a variation in the sampling frequency. Tapered grooves could be filled with extra fibers at increasing depth into the calorimeter – these varying bundles become difficult to calibrate. One other solution is to make extra holes from the back which terminate at stepped radii. This is also cumbersome, introduces constant terms and is not easy to calibrate.

By contrast, readout with plates, as in the present HE endcap calorimeter, allows relatively easy definition of pointing physics towers by adjusting the shape, area and placement of successive plates, and maintains constant sampling fraction at any angle (albeit with decreasing sampling frequency at lower $\eta$).

*3) Radiation Hardness:*

Radiation hardness is insufficient for plastic scintillating fiber calorimeters at SLHC, NLC/T-LEP, and beyond, and are almost impossible to replace, when interwoven with dual readout quartz fibers; by contrast, in a plate configuration, the plates in the first interaction lengths could be replaced, with proper design.:

a) Some scintillating plates, such as tiles of $BaF_2$, are sufficiently radiation resistant to be used in HE *$2.4 \leq \eta \leq 4$*. We have also shown that pTP and ZnO coated quartz plates can be scintillating rad-hard tiles, with sufficient raddam resistance to work in the first 1-2 $\lambda$.

b) With proper design, organic scintillating plates at the highest values of $\eta$ and covering the first 1-2 $\lambda$ can be replaced, or could be made compatible with replaceable liquids.

c) The WLS fibers can be designed to be replaceable, with straight through WLS fibers - see figure below - and to be coupled to radhard large clear quartz transport fibers.

d) In the plate configuration, the first $\lambda$ scintillating plates could be Liquid Scintillator.

e) Compact fiber calorimeters configurations with dual readout *require* photodetectors behind the calorimeter, vulnerable to backgrounds from punchthrough, and to raddam, whereas the photodetectors in dual readout plate calorimeters can be taken outside the calorimeter as in HE/HB. At present, no photodetector can work in the radiation and B-field environment <30cm from the beampipe, or behind only 10 $\lambda$.

*4) Flexibility:*
a) Longitudinal segmentation in a plate configuration allows direct e-m front ends (high-Z absorbers, finer sampling, followed by Cu or Fe absorbers with cruder sampleing etc) to more easily find the em/hadron fractions in an incident jet,while still compensated, with longitudinal analysis of events; and
b) The plate configuration allows lower index n materials in Cerenkov plates for better electromagnetic Cerenkov vs hadronic scintillation discrimination - such as silica aerogels or fluoride-based materials. Plates also enable different scintillators in front vs back, such as $BaF_2$ in front and scintillator a few Lambda back.

**Conclusions/Summary:** Rad-hard scintillating tiles with parallel replaceable WLS fiber readout is possible as an upgrade for the existing HE Endcap calorimeter absorber in CMS. The concept of combining Cerenkov compensated rad-hard tile calorimetry as upgradeable to Analog Particle Flow Calorimetry, as a replacement for the HE endcap calorimeter extending to η ≤4 is an interesting alternative. A simple sampling calorimeter about 30x30 cm, ½ λ sampling in Fe or Cu, 10 λ deep, with 20 quartz plates sandwiched to 20 plastic scintillator plates would provide a confirmation of MC predictions for a Cerenkov compensated dual plate readout calorimeter, and provide a basis for a particle flow design. Applications in $e^+e^-$ collider detectors and where precision calorimetry is necessary in high rate experiments are many.